\documentclass[twocolumn,showpacs,preprintnumbers,amsmath,amssymb]{revtex4}
\usepackage{bbm}
\usepackage{mathrsfs}
\usepackage{amsfonts}
\usepackage{tipa}
\usepackage{graphicx}
\usepackage{dcolumn}
\usepackage{bm}

\begin{document}

\title{Heavy quark symmetry in strong decays of P-wave heavy-light mesons}
\author{Bing Chen$^{1,2}$, Ling Yuan$^1$ and Ailin Zhang$^1$\footnote{Corresponding author:
zhangal@staff.shu.edu.cn}} \affiliation{$^1$Department of Physics,
Shanghai University, Shanghai 200444, China\\
$^2$School of Physics and Electrical Engineering, Anyang Normal University, Anyang 455000, China
}

%\date{\today}

\begin{abstract}
Heavy quark symmetry and its breaking in strong decays of P-wave heavy-light mesons are examined within the EHQ's method. The consistence of theory with experiments indicates that the breaking of heavy quark symmetry in the two-body strong decays of these mesons is not large. The relation between the EHQ's method and the $^3P_0$ model is investigated, and the phenomenological transition strengths $\mathcal {F}^{j_q,j'_q}_{j_h,l}(0)$ of the P-wave heavy-light mesons within the EHQ method are analytically derived within the $^3P_0$ model.
\end{abstract}
\pacs{11.30.Hv; 14.40.Lb; 14.40.Nd\\
Keywords: heavy quark symmetry, $^3P_0$ model, strong decay}
\maketitle

\section{Introduction}

Heavy quark symmetry and the physics behind it~\cite{hqeta,hqetb,hqetc} were observed in heavy-light systems $30$ years ago, heavy quark effective theory (HQET) has been established and successfully applied in many processes. The heavy quark symmetry in strong decays has also been investigated.

In Ref.~\cite{hqs1}, some consequences of the heavy quark symmetry for heavy-quark spectroscopy, and in particular
for strong decay widths, have been obtained. The obtained amplitudes for strong decays of heavy mesons were proportional
to sums of the products of four Clebsch-Gordan coefficients. Through a consideration of heavy quark symmetry, implications about the spectroscopy of the light degrees of freedom are discovered. However, only relationships between the heavy-quark systems involving given states of these degrees
of freedom could be predicted.

In Ref.~\cite{ehq}, a concise formula for two-body strong decay rates has been proposed based on heavy quark symmetry. In the formula, there is a $6-j$ symbol and a phenomenological transition strength $\mathcal {F}^{j_q,j'_q}_{j_h,l}(0)$. The $6-j$ symbol exhibits the heavy quark symmetry in the strong decays of heavy-light mesons. The transition strength $\mathcal {F}^{j_q,j'_q}_{j_h,l}(0)$ is a phenomenological parameter fitted by experimental data. To fix the $\mathcal {F}^{j_q,j'_q}_{j_h,l}(0)$, K mesons were treated as heavy-light systems in Ref.~\cite{ehq}.

The transition strength $\mathcal {F}^{j_q,j'_q}_{j_h,l}(0)$ can also be obtained in theory. In Ref.~\cite{CQM}, the transition strength was constructed from a relativistic chiral quark model. In Ref.~\cite{cyz}, the transition strength derived within the $^3P_0$ model was employed for the analysis, but the detail was omitted. In this paper, the derivation will be presented in detail.

In experiment, more and more highly excited heavy-light mesons (with one heavy $c$ or $b$ quark/anti-quark and one light $u$, $d$ or $s$ anti-quark/quark) have been observed~\cite{pdg}. Among the highly excited mesons, the $P-$wave mesons have often been concentrated on. In this paper, the $P-$wave heavy-light mesons with the following spin, parity $J^P$ and total angular momenta of the light degrees of freedom $j_l$: $0^+$ ($j_l=1/2$), $1^+$ ($j_l=1/2$), $1^+$ ($j_l=3/2$) and $2^+$ ($j_l=3/2$), will be denoted with $A^*_0$, $A^\prime_1$, $A_1$ and $A^*_2$, respectively. There is no mixing between $A^*_0$ and $A^*_2$, but there exists possible mixing between $A^\prime_1$ and $A_1$. When there exists mixing, it is more difficult to identify the states. How to identify the mixing of these states is also an important urgent topic.

Now let us have a glimpse at those $P-$wave heavy-light mesons. $D_0^\star(2400)$ and $D_{s0}^*(2317)^\pm$ are identified with the $A^*_0$ D and $D_s$, respectively. $B^*_0$ or $B^*_{s0}$ has not been observed. $D_2^\star(2460)$ and $D_{s2}^*(2573)^\pm$ are supposed the $A^*_2$ D and $D_s$, and $B^*_2(5747)$ and $B^*_{s2}(5840)^0$ are believed the $A^*_2$ B and $B_s$, respectively.

As for the $J^P=1^+$ axial vector mesons ($A'_1$, $A_1$ or the mixtures of them), $D_1(2420)^0$ was observed by Belle
collaboration~\cite{belle} but not confirmed by any other collaboration. $D_1(2430)^0$, $D_{s1}(2460)^\pm$ and $D_{s1}(2536)^\pm$
were observed by different collaborations~\cite{pdg}, where $D_1(2430)^0$ and $D_{s1}(2536)^\pm$ were supposed the $A_1$ D and $D_s$, respectively~\cite{pe}. $B_1(5721)$ and $B_{s1}(5830)$ were also observed by different collaborations~\cite{pdg}, but $B^*_J(5732)$ was only observed by $L3$ Collaboration~\cite{l3}.

The effects of heavy quark symmetry and its breaking in the mass spectrum and in the weak processes have been comprehensively explored. However, the heavy quark symmetry and its breaking in the strong decays of the highly excited heavy-light mesons have seldom been examined. With more and more data of the heavy-light mesons accumulated, it is possible to examine the heavy quark symmetry and its breaking in these mesons.

The paper is organized as follows. In Sec.II, the
two-body strong decays of $P-$wave heavy-light mesons are studied
to examine the heavy quark symmetry. In Sec.III, we present the detail of deriving the phenomenological transition strengths $\mathcal {F}^{j_q,j'_q}_{j_h,l}(0)$ of the P-wave heavy-light mesons within the $^3P_0$ model. Our
conclusions and discussions are included in Sec.IV.

\section{Heavy quark symmetry in the strong decays of heavy-light mesons}

For convenience, here and in what follows the notation $nL(J^P,
j_q)$ is used to label the heavy-light mesons. When an excited heavy-light meson $H$,
characterized by $nL(J^P, j_q)$, decays to a heavy-light meson
$H^\prime$ ($n'L'({J'}^{P'}, {j'}_q)$) and a light hadron $h$ with
spin $s_h$ and orbital angular momentum $l$ relative to $H'$, the
two-body strong decay width (the EHQ's formula) is~\cite{ehq}
\begin{eqnarray}\label{eq1}
\Gamma^{H\rightarrow H^\prime h}=\zeta (\mathcal
{C}^{s_Q,j'_q,J'}_{j_h,j_q,J})^2\mathcal
{F}^{j_q,j'_q}_{j_h,l}(0)p^{2l+1}\exp(-\frac{p^2}{6\beta^2}),
\end{eqnarray}
where
\begin{eqnarray*}
\mathcal {C}^{s_Q,j'_q,J'}_{j_h,J,j_q}=\sqrt{(2J'+1)(2j_q+1)}\left\{
           \begin{array}{ccc}
                    s_Q  & j'_q & J'\\
                    j_h  & J    & j_q\\
                    \end{array}
     \right\}
\end{eqnarray*}
is a normalized $6-j$ Wigner coefficient and $\overrightarrow{j}_h=\overrightarrow{s}_h+\overrightarrow{l}$. $\zeta$ is a flavor factor,
$\mathcal {F}^{j_q, j'_q}_{j_h,l}(0)$ is a phenomenological parameter (transition strength),
and $p$ is the momentum of decay products in the rest frame of $H$.
The $6-j$ symbol of the coefficients $\mathcal {C}$ exhibits the heavy
quark symmetry in the strong decays of heavy-light
mesons~\cite{hqs1,hqs2}. The allowed transitions between two doublets are governed by transition
strength $\mathcal {F}^{j_q,j'_q}_{j_h,l}(0)$.

The flavor factor $\zeta$ corresponding to some strong decay channels of D, $D_s$, B and $B_s$ mesons have been computed and presented in Table I.

\begin{table}
\begin{tabular*}{80mm}{c@{\extracolsep{\fill}}cccc}
\hline\hline
$X\rightarrow H'+h$ &      $\zeta$  \hspace {0.3cm}      &    $X\rightarrow H'+h$   &     $\zeta$  \\
\hline
$D^{0,\pm}\rightarrow D^{0,\pm}+\pi^0$    &    1 \hspace {0.3cm}     & $B^{0,\pm}\rightarrow B^{0,\pm}+\pi^0$       &    1 \\
$D^0(\overline{D}^0)\rightarrow D^\pm+\pi^\mp$    &    2 \hspace {0.3cm}     & $B^0(\overline{B}^0)\rightarrow B^\pm+\pi^\mp$     &    2 \\
$D^\pm\rightarrow D^0(\overline{D}^0)+\pi^\pm$    &    2 \hspace {0.3cm}     & $B^\pm\rightarrow B^0(\overline{B}^0)+\pi^\pm$     &    2 \\
$D^{0,\pm}\rightarrow D^{0,\pm}+\eta$    &    1/3\hspace {0.3cm}     & $B^{0,\pm}\rightarrow B^{0,\pm}+\eta$ &  1/3  \\
$D_s\rightarrow D^0+K$    &    2 \hspace {0.3cm}     & $B_s\rightarrow B^0+K$     &    2 \\
$D_s\rightarrow D^\pm+K^\mp$    &    2 \hspace {0.3cm}     & $B_s\rightarrow B^\pm+K^\mp$     &    2 \\
$D_s\rightarrow D^\pm+K^\mp$     &    4/3 \hspace {0.3cm}     & $B_s\rightarrow B_s+\eta$     &    4/3 \\
\hline\hline
\end{tabular*}
\caption{The flavor factor $\zeta$ corresponding to some strong decay channels of D, $D_s$, B and $B_s$ mesons.} \label{table 3}
\end{table}

The EHQ's method is based on the consideration of heavy quark symmetry, therefore any deviation of theoretical predictions from experiments will
provides an insight into the heavy quark symmetry and its breaking of the heavy-light mesons.

In the original Eq.~(\ref{eq1}), the transition strength was fixed by K mesons. In our computation, $D_0^*(2400)^0$ and
$D^\star_2(2460)^0$ are used as the experimental inputs. In HQET, mesons of $S$ doublet
decay to $H$ doublet and a light meson through an $s$-wave, mesons of the $T$ doublet decay to $H$ doublet and a light meson
through a $d$-wave. Accordingly, the transition strength $\mathcal
{F}^{\frac{1}{2},\frac{1}{2}}_{0,0}(0)$ ($s$-wave) and $\mathcal
{F}^{\frac{3}{2},\frac{1}{2}}_{2,2}(0)$ ($d$-wave) are fixed at
$0.267$ and $1.07$ GeV$^{-4}$, respectively.
In terms of these fixed transition strength, the decay widths of all possible $P-$wave heavy-light mesons~\cite{pdg} have been calculated and presented in the second column of Table II.

As shown in the next section, the transition strength $\mathcal
{F}^{\frac{1}{2},\frac{1}{2}}_{0,0}(0)$ ($s$-wave) and $\mathcal
{F}^{\frac{3}{2},\frac{1}{2}}_{2,2}(0)$ ($d$-wave) can also be obtained within the $^3P_0$ model. The corresponding decay widths of these $P-$wave heavy-light mesons are presented in the third column of Table II.

\begin{table}
\begin{tabular*}{85mm}{c@{\extracolsep{\fill}}ccc}
\toprule Width(MeV)  &     EHQ's         &  $^3P_0$  & Expt. \\
\hline
$D_0^*(2400)^0$   &  267 (input)   &  267 (input)        & $267\pm40$\\
$D_0^*(2400)^\pm$   &  298.3   &  229.1        & $283\pm40$\\
$D_1^*(2430)^0$   &  253.4   &  274.3     &   $384^{+130}_{-110}$  \\
\hline
$D_1(2420)^0$   & 16.0 & 16.0 &$27.1\pm2.7$\\
$D_1(2420)^\pm$   & 16.6 & 16.6 &$25\pm6$\\
$D^*_2(2460)^\pm$   & 50.2 &  50.2  & $37\pm6$ \\
$D^*_2(2460)^0$   &  49 (input)   &  49 (input)    &  $49.0\pm1.4$   \\
\hline\hline
$D_{s0}^*(2317)^\pm$&$0.010^{(B)}\hspace {0.05cm}0.128^{[E]}$&$0.014^{(B)}\hspace {0.05cm}0.174^{[E]}$& $<3.8$ \\
$D_{s1}^*(2460)^\pm$&$0.010^{(B)}\hspace {0.05cm}0.128^{[E]}$&$0.014^{(B)}\hspace {0.05cm}0.174^{[E]}$& $<3.5$ \\
\hline
$D_{s1}(2536)^\pm$& 0.43& 0.43 & $<2.3$\\
$D_{s2}^*(2573)^\pm$&   22.9    &   22.9  & $20\pm5$ \\
\hline\hline
 $B_0^*$   &     -     &    -      &       -     \\
 $^\dagger B_1^*(5670)$&      223.4    &      276.1    &    $70\pm46$   \\
\hline
$B_1(5721)^0$& 17.2 & 17.2 &       -     \\
$B_2^*(5747)^0$&  28.0   &   28.0  &    $23^{+5}_{-11}$ \\
\hline\hline
 $B_{s0}^*$&     -     &     -     &      -      \\
$B_{s1}^*$&      -    &      -    &       -     \\
\hline
$B_{s1}(5830)^0$& 0.026 & 0.026 &   narrow \\
$B_{s2}^*(5840)^0$&   1.62   &  1.62   &   narrow  \\
\hline\hline
\end{tabular*}
\caption{Decay widths of the $D$ $D_s$ $B$ and $B_s$ from theories and experiments. The upper right sign $(B)$ indicates the
isospin-forbidden decays through the $SU(2)$ breaking mechanism and
$(E)$ indicates the decays through electromagnetic interactions. Candidates marked
with slash are those omitted in PDG. $^\dag B_1^*(5670)$ has not been listed
in PDG.} \label{table 3}
\end{table}

The electromagnetic interaction and $SU(2)$ isospin symmetry breaking are believed to be two possible mechanisms for the isospin-violates decays of   $D_{s0}^*(2317)^\pm$ and $D_{s1}^*(2460)^\pm$~\cite{Azimov}. The decay widths given by these two ways are listed in Table II. For all these $P-$wave heavy-light mesons in Table II, it is easy to observe that theoretical predictions based on heavy quark symmetry agree well with experiments in most cases. For the decay width of $B_1^*(5670)$, the predicted result seems much larger than the experimental one.

As pointed in the introduction, $A'_1$ ($1P(1^+, \frac{1}{2}))$ and $A_1$ ($1P(1^+, \frac{3}{2})$) may mix with each other.
In fact, there exists evidence that the axial vector mesons $D_{s1}^*(2460)^\pm$ and
$D_{s1}(2536)^\pm$ are mixtures of $A'_1$ and $A_1$~\cite{pdg}. The $s-$channel decay of $D_{s1}(2536)^\pm$ has been observed and the measured
branching ratio~\cite{pdg}
\begin{eqnarray}
\frac{\Gamma((D^*(2010)^+ K^0)_{s-wave})}{\Gamma(D^*(2010)^+
K^0)}\approx0.72\pm0.05\pm 0.01
\end{eqnarray}

Therefore, it is very possible that
\begin{eqnarray}\label{eq3}
\begin{aligned}
 \left(
           \begin{array}{c}
                     D_{s1}^*(2460)^\pm\\
                     D_{s1}(2536)^\pm\\
                    \end{array}
     \right)&=\left(
           \begin{array}{cc}
                    cos\phi  & -sin\phi \\
                    sin\phi  & cos\phi\\
                    \end{array}
     \right)  \left(
           \begin{array}{c}
                     1P(1^+, \frac{1}{2}) \\
                     1P(1^+, \frac{3}{2}) \\
                    \end{array}
     \right)\\&=\left(
           \begin{array}{cc}
                    cos\varphi  & -sin\varphi \\
                    sin\varphi  & cos\varphi\\
                    \end{array}
     \right)  \left(
           \begin{array}{c}
                     1^3P_1\\
                     1^1P_1\\
                    \end{array}
     \right),
\end{aligned}
\end{eqnarray}
where
\begin{eqnarray*}
\varphi = \phi + 35.3^\circ.
\end{eqnarray*}

In existing literature, there is not yet a consensus on the mixing angle $\phi$ [or $\varphi$].
In Ref.~\cite{mixing}, the mixing angle $\phi$ are determined at $2.9^0\sim8.6^0$
and $0.9^0\sim2.6^0$ for D and B systems, respectively. In a
hadron loops model, the mixing angle was found at
$\varphi\simeq29^o\sim35^o$~\cite{Zhou}. When a
unitary rotation between the bases of $Q\overline{q}$
mesons $(J^2, j_q^2, s_Q^2, J_z)$ and $q\overline{q}$ mesons $(J^2,
L^2, S^2, J_z)$  is performed, $\phi\simeq0.3^o\sim6.3^o$~\cite{Ebert}.

In Refs.~\cite{cyz,ycz}, the mixing between the $2S$ and the $1D$ $D_s$ is
investigated and the mixing angle is fixed. In a similar way,
the mixing angle in Eq. (\ref{eq3}) can be determined as follows.

In our process, the mixing angle $\phi$ is treated as a free variable and fixed by the
comparison of theoretical prediction with experiments. In Fig. 1, the dependence of two branch
ratios, $\Gamma(D^*(2007)^0 K^+)/\Gamma(D^{*+} K^0)$ and
$\Gamma(D^{*+} K^0)_s/\Gamma(D^{*+} K^0)$, on the mixing
angle $\phi$ is presented. In plotting the red lines and the blue
lines, the predicted masses of pure $1^3P_1$ and $1^1P_1$
in Ref.~\cite{ig} and Ref.~\cite{li}, respectively, are employed and $\mathcal {F}^{j_q,j'_q}_{j_h,l}(0)$ is fixed by
experimental data.
The mixing angles determined by two different masses inputs are used as a
reasonable boundaries of the mixing angle. Obviously, $D_{s1}(2536)^\pm$ is $1P(1^+, \frac{3}{2})$ dominant. In Fig. 2, similar figures
are plotted with $\mathcal {F}^{j_q,j'_q}_{j_h,l}(0)$ obtained within the $^3P_0$ model.

\begin{figure}[ht]
\begin{center}
\includegraphics[width=8.4cm,keepaspectratio]{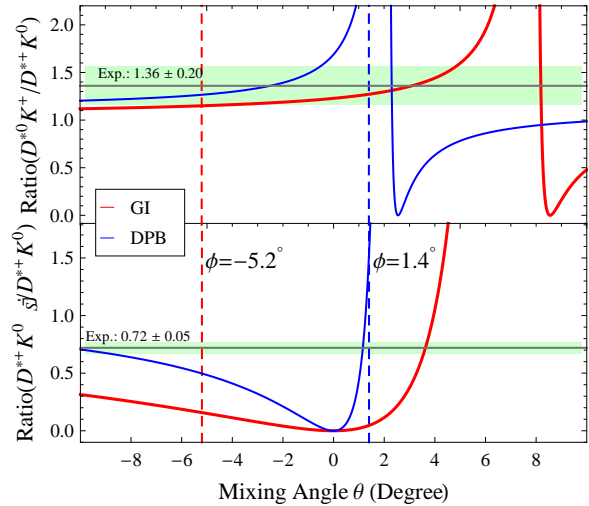}
\caption{$\Gamma(D^{*0}K^+)/\Gamma(D^{*+}K^0)$ and
$\Gamma(D^{*+}K^0)_s/\Gamma(D^{*+}K^0)$ versus the mixing angle
$\theta$.}
\end{center}
\end{figure}

\begin{figure}[ht]
\begin{center}
\includegraphics[width=8.4cm,keepaspectratio]{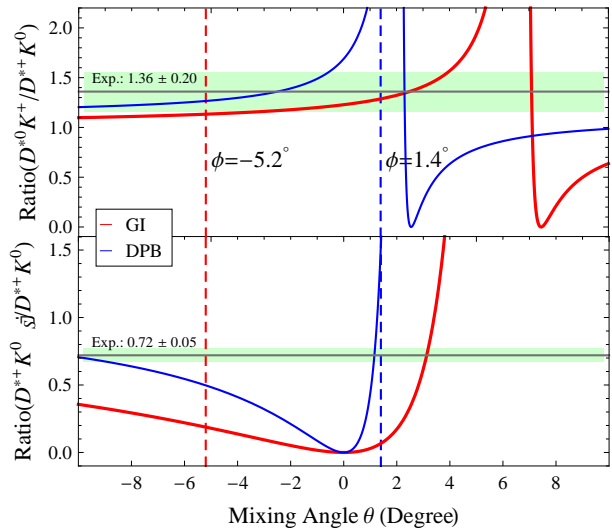}
\caption{The branching fractions
$\Gamma(D^{*0}K^+)/\Gamma(D^{*+}K^0)$ and
$\Gamma(D^{*+}K^0)_s/\Gamma(D^{*+}K^0)$ versus the mixing angle
$\theta$}
\end{center}
\end{figure}

In addition to the decay widths, some branching ratios of the heavy-light mesons have been measured. Accordingly, these ratios are calculated in terms of Eq. (\ref{eq1}). All the theoretical and experimental results are presented in Table III. Obviously, theoretical predictions of the branching ratios agree quite well with experiments.

In summary, theoretical prediction of the strong decays widths and branching ratios based on heavy quark symmetry is consistent with experiment, which indicates that the breaking of heavy quark symmetry in the two-body strong decays of these $D_s$ is not large.

\begin{table}
\renewcommand\arraystretch{1.8}
\begin{tabular*}{80mm}{c@{\extracolsep{\fill}}|ccc}
\toprule Branching ratios \hspace {0.3cm}  &     EHQ's  & Expt. \\
\hline
$\frac{\mathcal{B}(D^*(2010)^\pm\rightarrow D^0\pi^+)}{\mathcal{B}(D^*(2010)^\pm\rightarrow D^+\pi^0)}$ \hspace {0.3cm}    &  2.21      &   $2.21\pm0.05$  \\
$\frac{\mathcal{B}(D_2^*(2460)^0\rightarrow D^+\pi^-)}{\mathcal{B}(D_2^*(2460)^0\rightarrow {D^\star+}\pi^-)}$   \hspace {0.3cm}  &  2.21      &   $1.56\pm0.16$  \\
$\frac{\mathcal{B}(D_2^*(2460)^\pm\rightarrow D^0\pi^+)}{\mathcal{B}(D_2^*(2460)^\pm\rightarrow D^{\star+}\pi^0)}$   \hspace {0.3cm}  &  2.22      &   $1.90\pm1.10$  \\
$\frac{\mathcal{B}(D_{s2}^*(2573)^+\rightarrow D^{\star0}K^+}{\mathcal{B}(D_{s2}^*(2573)^+\rightarrow D^0K^+)}$   \hspace {0.3cm}  &  0.10      &   $<0.33$  \\
$\frac{\mathcal{B}(B_2^*(5747)^0\rightarrow B^{\star+}\pi^0)}{\mathcal{B}(B_2^*(5747)^0\rightarrow B^+\pi^0)}$    \hspace {0.3cm} &  0.90       &  $1.10\pm0.50$  \\
\hline\hline
\end{tabular*}
\caption{The branching ratios from EHQ's formula and experiments~\cite{pdg}.} \label{table 3}
\end{table}

\section{Transition strength
$\mathcal {F}^{j_q, j'_q}_{j_h,l}(0)$ in the $^3P_0$ model}

The $^3P_0$ model, also known as the quark-pair creation model, was
originally introduced by Micu~\cite{Micu} and further developed by
Le Yaouanc \emph{et al}.,~\cite{Yao1}. It has been extensively employed
to study the strong decay of hadrons. The key idea of the $^3P_0$ decay model is
the assumption of the creation of a
$0^{++}(^3P_0)$ quark-antiquark pair from the vacuum~\cite{Micu}.

The interaction Hamiltonian $H_I$ relevant to the pair production involves two Dirac quark fields~\cite{Ackleh}
\begin{eqnarray}\label{eq5}
H_I^{^3P_0}=\gamma\int
d^3k\overline{\psi}_f(k)\psi_i(k),
\end{eqnarray}
where $m_q$ is the mass of the produced quarks. The dimensionless constant $\gamma$ is the
$^3P_0$ pair production coupling constant which can be extracted by
fitting the data, where a color matrix element
$\langle\omega_B\omega_C\mid\omega_A\omega_0\rangle=1/3$ is included.

In a non-relativistic quark model, the helicity decay amplitude for a process $A\to BC$ could be
written as~\cite{Yao2}
\begin{equation}\label{helicity}
\begin{aligned}
\mathcal {M}^{helicity}&=\langle B C\mid H_I^{^3P_0}\mid
A\rangle\\
&=\sum\limits_{l_i,s_i}\mathcal
{W}_{CG}(J_i,L_i,S_i;j_i,l_i,s_i)\mathcal {I}_{l_i}(p),
\end{aligned}
\end{equation}

In the center-of-mass system, the spatial overlap integral is
\begin{equation}
\begin{aligned}
\mathcal {I}_{l_i}(p)= \int d^3k \psi^*_{n_B,L_B,l_B}
\psi^*_{n_C,L_C,l_C} \psi_{n_A,L_A,l_A}\mathcal {Y}_{1l_v}.
\end{aligned}
\end{equation}

The partial width is
\begin{equation}\label{eq7}
\begin{aligned}
\Gamma_{A\rightarrow B C}&=2\pi \zeta\frac{\widetilde{M}_B\widetilde{M}_C}{\widetilde{M}_A}p\\
&\times \sum\limits_{L,S}\mid\langle J_A,j_A\mid L,0; S,j_A\rangle\\
&\hspace {0.9cm}\langle J_B,j_B; J_C,j_C\mid S,j_A \rangle\sum\limits_{l_i,s_i}\mathcal {M}^{helicity}\mid^2\\
&=2\pi\zeta
\frac{\widetilde{M}_B\widetilde{M}_C}{\widetilde{M}_A}p\sum\limits_{LS}\mid
\mathcal {M}_{LS}\mid^2,
\end{aligned}
\end{equation}
where the mock-meson masses $\widetilde{M}_i$ are explained in detail in
Ref.~\cite{Isgur}. $\zeta$ is an overlap integral of the flavor wave function
\begin{eqnarray*}
\zeta = \langle\varphi_B \varphi_C\mid \varphi_A\varphi_0\rangle
\end{eqnarray*}
where $\varphi_0$ is a
flavor singlet of vacuum.

Due to the heavy quark symmetry in the heavy-light systems, the partial wave amplitude in the base $(J^2, j_q^2, s_Q^2, J_z)$ can
also be written directly as~\cite{hqs1,ehq}
\begin{equation}
\begin{aligned}
\mathcal {M_{LS}}&=(-1)^{s_Q+j_h+J'+j_q}\mathcal
{C}^{s_Q,j'_q,J'}_{j_h,j_q,J}\mathcal {A}_R(j_h,l,j_q,j'_q),
\end{aligned}
\end{equation}
where the reduced amplitude $\mathcal {A}_R$ is
\begin{eqnarray*}
\delta_{s'_{Qz},s_{Qz}}\langle
h(j_h,j_{hz});\frac{1}{2},s'_{Qz};j'_q,j'_{qz}\mid H_I^{^3P_0}\mid
\frac{1}{2},s_{Qz};j_q,j_{qz}\rangle.
\end{eqnarray*}

When the simple harmonic oscillator approximation is employed as
the meson space wave functions, the helicity amplitude in Eq.~(\ref{helicity}) is
obtained analytically in the base $(J^2, L^2, S^2, J_z)$. Therefore, a unitary rotation between the base
$(J^2, j_q^2, s_Q^2, J_z)$ and $(J^2, L^2, S^2, J_z)$ is usually employed as follows~\cite{cahn,Ebert}
\begin{equation}\label{rotation}
\begin{aligned}
\mid J;
j_q\rangle=&(-1)^{J+L+1}\sqrt{(2S+1)(2j_q+1)}\\
   &\times\left\{
           \begin{array}{ccc}
                    1/2  & 1/2 & S\\
                    L    & J   & j_q\\
                    \end{array}
     \right\}\mid J; S\rangle.
\end{aligned}
\end{equation}

In this way, the transition amplitude $\mathcal {M}_{LS}$ of P-wave mesons have been calculated in the
base $(J^2, L^2, S^2, J_z)$. The results are presented in Table IV
with
\begin{eqnarray*}
&f_S=\frac{2^{9/2}}{3^2}(1-\frac{2}{9}x^2)\\
&f_P=\frac{2^5}{3^{5/2}}x \hspace {0.5cm}
f_D=-\frac{2^{11/2}}{3^4}x^2.
\end{eqnarray*}

\begin{table}
\begin{tabular*}{85mm}{c@{\extracolsep{\fill}}cc}
\toprule $X$($n^{2s+1}L_J$)\hspace {0.2cm}  &      $H(1^1S_0)+h(1^1S_0)$  \hspace {0.2cm}     &    $H(1^3S_1)+h(1^1S_0)$    \\
\hline
$1^3S_1$  \hspace {0.2cm}&    $-\sqrt{1/3}f_P$    \hspace {0.2cm} &   $\sqrt{2/3}f_P$  \\
\hline
$1^3P_0$  \hspace {0.2cm}&    $f_S$    \hspace {0.2cm} &   $-$  \\
$1^3P_1$  \hspace {0.2cm}&   $-$    \hspace {0.2cm} &   $\sqrt{2/3}f_S$ \hspace {0.3cm} $\sqrt{1/3}f_D$ \\
$1^1P_1$  \hspace {0.2cm}&   $-$    \hspace {0.2cm} &   $-\sqrt{1/3}f_S$ \hspace {0.2cm} $\sqrt{2/3}f_D$ \\
$1^3P_2$  \hspace {0.2cm}&   $-\sqrt{2/5}f_D$    \hspace {0.2cm} &  $\sqrt{3/5}f_D$ \\
\hline\hline
\end{tabular*}
\caption{The decay amplitudes $\mathcal {M}_{LS}$ where a
coefficient $\frac{\gamma}{\pi^{1/4}\beta^{1/2}}e^{-x^2/12}$ has
been omitted~\cite{Ackleh}.} \label{table3}
\end{table}

In terms of Eq.~(\ref{rotation}), one can get the
reduced amplitude $\mathcal {A}_R$ easily. Through a comparison of Eq.~(\ref{eq1}) with Eq.~(\ref{eq7}), the transition strength
$\mathcal {F}^{j_q,j'_q}_{j_h,l}(0)$ are finally obtained
\begin{eqnarray}\label{eq11}
\mathcal {F}^{\frac{1}{2},\frac{1}{2}}_{0,0}(0) =\mathcal
{G}\frac{1}{\beta}(1-\frac{2}{9}\frac{p^2}{\beta^2})^2 \hspace {0.5cm} \mathrm{for\hspace {0.2cm} S-wave}\\
\mathcal{F}^{\frac{1}{2},\frac{1}{2}}_{1,1}(0) =\mathcal
{G}\frac{2}{3}\frac{1}{\beta^3}\hspace {1.52cm} \mathrm{for\hspace {0.2cm} P-wave} \\
\mathcal{F}^{\frac{3}{2},\frac{1}{2}}_{2,2}(0) =\mathcal
{G}\frac{2^2}{3^4}\frac{1}{\beta^5}\hspace {1.34cm}\mathrm{for\hspace
{0.2cm} D-wave}
\end{eqnarray}
where the constant
\begin{eqnarray*}
\mathcal
{G}=\pi^{1/2}\gamma^2\frac{2^{10}}{3^4}\frac{\widetilde{M}_B\widetilde{M}_C}{\widetilde{M}_A}.
\end{eqnarray*}

$\mathcal {F}^{\frac{1}{2},\frac{1}{2}}_{0,0}(0)$ depends on both $p$ and $\beta$, $\mathcal{F}^{\frac{1}{2},\frac{1}{2}}_{1,1}(0)$ and $\mathcal{F}^{\frac{3}{2},\frac{1}{2}}_{2,2}(0)$ depend only on $\beta$. Other $\mathcal {F}^{j_q,j'_q}_{j_h,l}(0)$ can be obtained in the same way.

Within the $^3P_0$ model, $\beta$ was usually fixed by the data of many light mesons. The reasonable value of $\beta$ is often supposed around $0.35-0.4$ GeV~\cite{Ackleh}.

In the reasonable region of $\beta$, $\mathcal {F}^{j_q,j'_q}_{j_h,l}(0)$ fixed by experiments are consistent with $\mathcal {F}^{j_q,j'_q}_{j_h,l}(0)$ obtained within the $^3P_0$ model. If $\beta=0.38$ GeV is employed as in Ref.~\cite{cyz}, the following ratios of transition strengths are obtained
\begin{eqnarray*}
\frac{\mathcal {F}^{\frac{1}{2},\frac{1}{2}}_{0,0}(0)}{\mathcal
{F}^{\frac{3}{2},\frac{1}{2}}_{2,2}(0)}\hspace
{1.4cm}0.30(\mathrm{EHQ's})\hspace {0.8cm}0.33(^3P_0)\\
\frac{\mathcal {F}^{\frac{1}{2},\frac{1}{2}}_{1,1}(0)}{\mathcal
{F}^{\frac{3}{2},\frac{1}{2}}_{2,2}(0)}\hspace
{1.4cm}1.64(\mathrm{EHQ's})\hspace {0.8cm}1.95(^3P_0)
\end{eqnarray*}

Obviously, the ratios of the transition strength obtained within the $^3P_0$
model are consistent with those fixed by experimental
data (EHQ's method) within uncertainties.

\section{CONCLUSIONS AND DISCUSSIONS}
In this work, the formula (Eq. (\ref{eq1})) proposed by Eichten {\it et al.} is employed to compute the strong decay of P-wave heavy-light mesons. Theoretical prediction of the decay widths and branching ratios is consistent with experiments. As well known, Eq. (\ref{eq1}) is based on the
consideration of heavy quark symmetry. Our results indicate that the heavy quark symmetry keeps well in the $P-$wave heavy-light mesons.

The axial vector mesons $D_{s1}^*(2460)^\pm$ and
$D_{s1}(2536)^\pm$ are mixtures of $A'_1$ and $A_1$ as indicated in Eq. (\ref{eq3}).
The fitted mixing angle $\phi$ is around $-5.2^\circ-1.4^\circ$.

In the original work of Eichten {\it et al.}, the phenomenological transition strength $\mathcal {F}^{j_q, j'_q}_{j_h,l}(0)$
is fitted by experiments. In this paper, the parameters $\mathcal {F}^{j_q, j'_q}_{j_h,l}(0)$ of P-wave heavy-light mesons are analytically obtained within the $^3P_0$ model. When the universal width parameter $\beta$ is chosen with $0.38$ GeV, the predicted results agree well with those fitted by experiments.

Of course, the $1/m_Q$ corrections in the strong decays has not been taken into account in the method of Eichten {\it et al.}. How large these corrections are is still not clear. For lack of data, the heavy quark symmetry in strong decays of other higher excited heavy-light mesons have not been examined.

\begin{acknowledgments}
Bing Chen thanks Mr. Zhi-Feng Sun and Prof. Xiang Liu very much for useful
communication. This project is supported by the
National Natural Science Foundation of China (NSFC) under grant
No. 11075102. Bing Chen is also supported by
Shanghai University under the Graduates Innovation Fund SHUCX111023 (No. A. 16-0101-10-018).
\end{acknowledgments}

\end{document}